\documentclass[12pt]{iopart}

\def\mstone{\wt{M}_{t1}}
\def\sn2w{\sin^2\theta_W^{}}
\def\bsgg{b \to s \gamma g}
\def\BSgamma{B \to X_s \gamma}

\def\HC{H^\pm}
\def\w{\omega}
\def\wt{\widetilde}

\def\mgr{m_{3/2}}
\def\acp{{\cal A}_{CP}}
\def\bsgamma{b \to s \gamma}

\def\to{\rightarrow}
\def\nn{\nonumber}
\def\gsim{~{\rlap{\lower 3.5pt\hbox{$\mathchar\sim$}}\raise
1pt\hbox{$>$}}\,}
\def\lsim{~{\rlap{\lower 3.5pt\hbox{$\mathchar\sim$}}\raise
1pt\hbox{$<$}}\,}

\def\PRD#1#2#3{19#2 {\it Phys. Rev.} {\bf D#1}, #3 }
\def\NPB#1#2#3{19#2 {\it Nucl. Phys.} {\bf B#1}, #3 }

\def\PLB#1#2#3{19#2 {\it Phys. Lett.} {\bf B#1}, #3 }
\def\PRL#1#2#3{19#2 {\it Phys. Rev. Lett.} {\bf #1}, #3 }

\begin{document}

\title[Author guidelines for IOPP journals]
{\Huge \bf Decay rate asymmetry
for $B \to X_s \gamma$ in SUSY model }

\author{
Mayumi Aoki
}

\address{Theory Group, KEK, Tsukuba, Ibaraki 305-0801, Japan }

\begin{abstract}

We discuss a rate asymmetry for the radiative $B$-meson decay $\BSgamma$
within the framework of the supersymmetric standard model based on $N=1$
supergravity.
This model contains
new sources of CP violation.
In spite of severe experimental constraints on the electric
dipole moment of the neutron,
a new CP-violating phase may not be suppressed, leading to
a sizable enhancement of the decay rate asymmetry.
The magnitude of the asymmetry is predicted to be larger than
that by the standard model
in wide parameter regions where the branching ratio
is consistent with its experimental bounds.
A possible maximal magnitude is about 0.1, which will be
well accessible at B factories.

\end{abstract}


\maketitle

     The radiative $B$-meson decay $\BSgamma$ is
sensitive to new physics around the electroweak scale.
Owing to a large mass of the $b$ quark, its inclusive mode is
well described by the free $b$-quark decays $\bsgamma$ and $\bsgg$,
to which new interactions may give sizable contributions.
In particular,  the supersymmetric standard model (SSM) has
various new sources for the contributions \cite{oshimo}.
Although the experimental results for the branching ratio \cite{cleo,aleph}
is consistent with the standard model (SM),
a decay rate asymmetry
\begin{eqnarray}
\acp = \frac{
   \Gamma(\overline{B}\to X_s\gamma)-\Gamma(B\to X_{\bar s}\gamma)}
   {\Gamma(\overline{B}\to X_s\gamma)+\Gamma(B\to X_{\bar s}\gamma)}
\label{eq:Acp}  \nn
\end{eqnarray}
could nontrivially differ in the SSM \cite{aoki} from the SM prediction,
which we discuss in this report.

     We assume the SSM based on $N=1$ supergravity and grand unification.
At the electroweak scale, the squarks of
interaction eigenstates are mixed in generation space.
The squarks, as well as the quarks, mediate
the processes of flavor-changing neutral current (FCNC).
The generation mixings among the up-type squarks are approximately
lifted by using the same matrices that diagonalize the mass
matrix of the up-type quarks.
As a result, the generation mixings in
the interactions between up-type squarks and down-type quarks
are described by the Cabibbo-Kobayashi-Maskawa
matrix for the quarks.

     The squark mass-squared matrices and the chargino
mass matrix have new sources of CP violation.
For physical complex parameters,
without loss of generality, we can take the dimensionless coefficient $A$
and the Higgsino mass parameter $\mu$, which are expressed as
$A = |A| \exp(i\alpha)$ and  $\mu = |\mu| \exp(i\theta)$, respectively.
We neglect the differences of the dimensionless coefficients
among flavors.
The CP-violating phase $\theta$ is severely
constrained by the experimental bounds on the electric
dipole moment (EDM) of the neutron \cite{kizukuri}.
If $\theta$ is of order unity,
the squarks are not allowed to have masses smaller than 1 TeV.
On the other hand, for a sufficiently small magnitude of $\theta$
and relatively large masses for gauginos, the squarks
can be of order 100 GeV with another CP-violating
phase $\alpha$ being unsuppressed \cite{aoki2}.
In this parameter region the SSM
may induce sizable CP violation without causing discrepancies for the EDM.

     The effective Hamiltonian for the decays $\bsgamma$ and $\bsgg$
is written generally by
\begin{eqnarray}
H_{eff} = -\frac{4G_F}{\sqrt 2} V_{ts}^*V_{tb}\sum_{j=1}^8
C_j(\mu) O_j(\mu),  \nn
\end{eqnarray}
\begin{center}
$ O_2 = \overline{s_L}\gamma_\mu c_L\overline{c_L}\gamma^\mu b_L, \quad
O_7 = \frac{e}{16\pi^2}m_b\overline{s_L}\sigma^{\mu\nu}b_R F_{\mu\nu}, \quad
O_8 = \frac{g_s}{16\pi^2}m_b\overline{s_L}\sigma^{\mu\nu}T^ab_R
             G_{\mu\nu}^a.  $
\end{center}
Here $C_j(\mu)$ denotes a Wilson coefficient evaluated at the scale $\mu$,
and $F_{\mu\nu}$ and $G_{\mu\nu}$ respectively represent the electromagnetic
and strong field strength tensors, $T^a$ being the generator of SU(3).
The contributions from the operators other than $O_2$, $O_7$, and $O_8$
are negligible.

     The Wilson coefficients $C_7$ and $C_8$ of the electroweak scale
receive contributions at the one-loop level.  Sizable new contributions
come from the diagrams in which the charginos $\w$ or the charged
Higgs bosons $\HC$ are exchanged together with the up-type squarks
or the up-type quarks, respectively.
The coefficients are expressed at the leading order as
$C_2(M_W) =1$ and
$C_j(M_W) = C_j^W(M_W)+C_j^{\HC}(M_W)+C_j^\w(M_W)$ $(j=7,8)$,
where $C_j^W$, $C_j^{\HC}$, and $C_j^\w$ stand respectively
for the contributions from $W$ bosons, charged Higgs bosons, and charginos.
In our scheme for the SSM, the one-loop diagrams mediated by the gluinos or
the
neutralinos with the down-type squarks cause only small effects
on both FCNC and CP violation.

     The contributions $C_7^\w(M_W)$ and $C_8^\w(M_W)$ have complex values,
owing to the physical complex phase intrinsic in the SSM.
Therefore, CP invariance is violated in the decay $\BSgamma$.
Another important effect by the SSM is that
$C_7^\w(M_W)$ and $C_8^\w(M_W)$ can be added to $C_7^W(M_W)$ and
$C_8^W(M_W)$ both constructively and destructively depending on
the parameter values.
On the other hand, $C_7^{\HC}(M_W)$ and $C_8^{\HC}(M_W)$ are added
constructively.
These effects make it possible to have a large magnitude for
the decay rate asymmetry while keeping the branching ratio
comparable with the SM value.

     The rate asymmetry in the decay $\BSgamma$ is attributed to
decay rate asymmetries for $\bsgamma$ and $\bsgg$, which are
induced by the interferences between
the tree level processes and the one-loop level processes with
absorptive parts.  The asymmetry $\acp$ is expressed in terms
of the Wilson coefficients at the $b$-quark mass scale \cite{kagan, greu}, 
which are obtained by solving the renormalization-group equations.

     The parameter values of the SSM are constrained by the measured
branching ratio of $\BSgamma$,
as well as direct searches for supersymmetric particles.
We calculate the decay width of $\BSgamma$ by
using the matrix elements and
anomalous dimensions at the next-to-leading order (NLO).
However, NLO corrections for the matching conditions of
$C_7^\w$ and $C_8^\w$ at the electroweak scale have not yet
obtained in general form.  Therefore,
the NLO matching conditions are taken into account only
for the W-boson contributions.

     The decay rate asymmetry for $B \to X_s \gamma$
is analyzed numerically together with its branching ratio.
Taking into account experimental constraints on the EDM of the neutron,
we assume $\alpha \sim \pi/4$ and $\theta \sim 0$ for the  CP-violating
phases,
and $\wt m_2 \gsim 500$ GeV for the SU(2) gaugino mass.
The mass parameters for the charged Higgs boson and the Higgsino
are taken for $M_{H^\pm}\sim 100$ GeV and $\mu \sim 100$ GeV,
respectively.  For simplicity, we assume that
the soft supersymmetry-breaking masses of the $t$ squarks
at the electroweak scale are given as $\wt m^2-cm_t^2$ and $\wt m^2-2cm_t^2$
respectively for $\wt t_L$ and $\wt t_R$.
A dimensionless constant $c$ is introduced to parametrize radiative
corrections to the squark masses though Yukawa interactions, with
$c=0.1-1$.
We take $\wt m /|A|m_{3/2} =0.5-2$ and $|A|m_{3/2} \lsim$ 1 TeV,
where $\mgr$ denotes the gravitino mass.
The ratio of the vacuum expectation values is considered to be
$\tan\beta=2-35$.
The energy cutoff parameter $\delta$ for the photon
is taken to be 0.99, though the decay rate
asymmetry is not changed so much by the choice of its value.

     For $\tan\beta= 10$,
the branching ratio lies within the experimental bounds
in the mass range 100 GeV$\lsim \mstone \lsim $400 GeV for the lighter
$t$-squark, where
the asymmetry has a value $0.02 \lsim |\acp| \lsim 0.07$.
The magnitude of the asymmetry becomes maximum at $\mstone \simeq 200$ GeV,
which gives roughly a minimum value of the branching ratio.
The maximal value of the asymmetry does not depend
significantly on $\tan \beta$.
The peaks of the asymmetry and the branching ratio are
roughly at the same value of $\mstone$, which increases with $\tan \beta $.

     Assuming that the CP-violating phase $\alpha$ is not suppressed
and the charged Higgs boson mass is of order 100 GeV,
the magnitude of the asymmetry is larger than 0.01 in wide parameter regions
consistent with the experiments on the branching ratio.
The SM prediction is smaller than 0.01.
If the charged Higgs boson is not heavy, the sum of the contributions
of $W$ and $\HC$ alone makes the decay width too large.
Therefore, in the SSM parameter region allowed by the branching ratio,
the chargino contribution has to be comparable with the contributions
of $W$ and $\HC$.  Then, a large CP asymmetry is induced.
If the charged Higgs boson is sufficiently heavy, its contribution
to the decay width is negligible.  Still, there are parameter regions
where the asymmetry is larger than 0.01 without conflicting with
the measured branching ratio.

     We have discussed the decay rate asymmetry for
the radiative $B$-meson decay $\BSgamma$ in the SSM based on $N=1$
supergravity and grand unification.  This model predicts various new
contributions to
the decay, among which the chargino
and the charged Higgs boson loop diagrams yield sizable effects.
Assuming an unsuppressed CP-violating phase intrinsic in the SSM,
the asymmetry can have a large value, maximally of order 0.1,
which is not expected by the SM.
Such a magnitude of the asymmetry
may well be detectable at B factories.  In particular,
the obtained experimental results that the branching ratio is
consistent with the SM would imply a large asymmetry.
The decay rate asymmetry of $\BSgamma$ seems worth
measuring in the present or near-future experiments.


\end{document}